# Extensive Divergence of Transcription Factor Binding in *Drosophila* Embryos with Highly Conserved Gene Expression

**Mathilde Paris[1]\*, Tommy Kaplan[1,2], Xiao Yong Li[1,3], Jacqueline E. Villalta[2], Susan E. Lott[1,4], Michael B. Eisen[1,2,3]\***

1 Department of Molecular and Cell Biology, University of California Berkeley, Berkeley, California, United States of America, 2 School of Computer Science and Engineering, The Hebrew University, Jerusalem, Israel, 3 Howard Hughes Medical Institute, University of California Berkeley, Berkeley, California, United States of America, 4 Department of Evolution and Ecology, University of California, Davis, California, United States of America

## Abstract

To better characterize how variation in regulatory sequences drives divergence in gene expression, we undertook a systematic study of transcription factor binding and gene expression in blastoderm embryos of four species, which sample much of the diversity in the 40 million-year old genus *Drosophila*: *D. melanogaster*, *D. yakuba*, *D. pseudoobscura* and *D. virilis*. We compared gene expression, measured by mRNA-seq, to the genome-wide binding, measured by ChIP-seq, for four transcription factors involved in early anterior-posterior patterning. We found that mRNA levels are much better conserved than individual transcription factor binding events, and that changes in a gene's expression were poorly explained by changes in adjacent transcription factor binding. However, highly bound sites, sites in regions bound by multiple factors and sites near genes are conserved more frequently than other binding, suggesting that a considerable amount of transcription factor binding is weakly or non-functional and not subject to purifying selection.

**Funding:** This work was funded by a Howard Hughes Medical Institute investigator award to MBE and by NIH grant HG002779 to MBE. MP was funded by a long-term post-doctoral fellowship awarded by the Human Frontier Science Program organization (HFSP) and by a Genentech Foundation fellowship. TK was funded by a Long-Term Post-Doctoral Fellowship by the European Molecular Biology Organization (EMBO), and is a member of the Israeli Center of Excellence (I-CORE) for Gene Regulation in Complex Human Diseases (no. 41/11), and the Israeli Center of Excellence (I-CORE) for Chromatin and RNA in Gene Regulation (1796/12). SEL was funded by NIH/NIGMS grant K99/R00-GM098448. The funders had no role in study design, data collection and analysis, decision to publish, or preparation of the manuscript.

**Competing Interests:** I have read the journal's policy and have the following conflicts: MBE is co-founder and member of the Board of Directors of PLOS.

* E-mail: thildeparis@gmail.com (MP); mbeisen@gmail.com (MBE)

## Introduction

In the pursuit of the genetic basis of phenotypic evolution, researchers have favored divergence of gene expression as a major source of diversity [1,2]. Changes in gene expression during animal development can have many origins, but it is widely assumed that the divergence of the enhancer sequences that drive complex spatial and temporal patterns of gene expression during development will play an important role in the evolution of morphology.

Enhancer sequences often undergo rapid changes during evolution [3–6], and the binding of transcription factors (TFs) to these and other sequences has been shown to be highly divergent in related species, where, based on functional conservation, little divergence was expected [7–14]. While many studies have compared gene expression between species at the genomic scale and reported various degrees of divergence (e.g. [15–17]), the downstream effects of TF binding divergence on gene expression has received relatively little attention [18].

One limitation of many of these studies is that significant changes in DNA sequence are often accompanied by extensive changes in morphology or physiology, complicating direct comparison of molecular phenotypes like transcription factor binding and gene expression. This is not, however, the case in *Drosophila* embryogenesis, which is highly conserved across the 40 million-year old *Drosophila* genus despite accumulating sequence changes equivalent to those separating different classes of amniotes [19].

To take advantage of this morphological conservation, we measured gene expression and the genome-wide binding of four TFs, namely Bicoid (BCD), Giant (GT), Hunchback (HB) and Krüppel (KR), involved in anterior-posterior (A-P) patterning in blastoderm embryos of four fully-sequenced species - *D. melanogaster*, *D. yakuba*, *D. pseudoobscura* and *D. virilis* – that span the genus, with divergence times between 5 and 40 million years ago [20] (Figure 1).

## Results

### Chromatin immunoprecipitation (ChIP) from divergent *Drosophila* species

We established large populations of *D. melanogaster* (Oregon R), *D. pseudoobscura* (MV2-25) and *D. virilis* (V46) for embryo collections. In our studies of transcription factor binding in *D. melanogaster* we generally use embryos from one hour collections aged for an additional two hours to target the cellular blastoderm

### Author Summary

Inter-species differences in gene expression during development are a major source of phenotypic diversity, yet the molecular origins of such differences are poorly understood. In this study we use a combination of biochemical and genomic methods to explore how an important component of the machinery of gene regulation varies between species. We mapped the binding sites bound by four specific proteins regulating gene expression along the head to tail axis of young embryos of four diverse species of fruit flies. We were surprised at the extent of variation we observed, especially as we increasing variation in gene expression between these same species' embryos. We conclude, based on various analyses of our binding divergence data, that most of the time these regulators bind to DNA they have no effect on gene expression, and therefore natural selection does not act to preserve these interactions between species.

stage [13,21,22], during which many key events in patterning occur. Because developmental timing varies between species, we optimized collection conditions for each species to obtain similar stage distributions (Table S1).

We fixed embryos in 1% formaldehyde to cross-link proteins to DNA, and purified chromatin. We immunoprecipitated cross-linked chromatin from all three species using rabbit polyclonal antibodies raised against *D. melanogaster* Bicoid (BCD), Hunchback (HB), Giant (GT) and Krüppel (KR) [3–6,22]. For each factor, we performed parallel ChIP using antibodies that were purified against recombinant versions of the corresponding *D. melanogaster* or *D. virilis* proteins (we were unable to affinity purify GT antisera against *D. virilis* GT). We used *D. virilis* proteins to avoid biases due to the recognition of a greater number of epitopes in *D. melanogaster* than in the more distantly related *D. pseudoobscura* and *D. virilis*. Following immunoprecipitation (IP), we sequenced the recovered DNA as well as input controls.

We generated a total of 20 ChIP datasets to which we added eight datasets from our previously published comparison of the binding of these factors in *D. melanogaster* and the closely related *D. yakuba* ([13], see Figure S1). We mapped reads to the corresponding genome sequences with Bowtie [23] (mapping statistics given in Table S2), and identified genomic regions significantly bound by each factor in any of the four species using two peak callers, Grizzly Peak [24,25] and MACS [26]. Roughly similar numbers of peaks were found in each ChIP (Table S3).

To minimize the effect of sample handling variation on the data, several of our ChIPs were carried out on pooled samples containing chromatin from multiple species, so that the IP was carried out in identical conditions. We could unambiguously assign more than 99% of reads to a species after sequencing, and verified that the small fraction of reads that we could not assign with confidence to a specific genome had almost no effect on peak height measurements (Figure S2). Replicates showed good reproducibility, especially for GT and HB IPs (Figure S3). For downstream analyses, we used log-transformed peak height values.

## Extensive divergence of transcription factor binding

To compare binding across species, we first aligned the genomes of all four species using Mercator [27] and Pecan [28] and identified orthologous regions present in all four genomes. We then projected the normalized binding intensities of bound regions from all IPs onto the coordinates of the whole-genome alignment and compared occupancy, as illustrated for the *even-skipped* locus (Figure 2). We obtained 2061 sets of orthologous regions bound by BCD, 4191 by GT, 4986 by HB and 5309 by KR. We also collapsed the 16,547 regions identified as bound by individual transcription factors, and merged them into a common set of 10,137 merged regions, bound by at least one factor in at least one species (see Figure 2). We then computed the occupancy of every factor in each species along each of these regions and compared values between species. Pairwise comparisons showed that binding intensity varies extensively between species (Figure 3A and Figure S4). As expected, the extent of conservation of sites for each factor decreased with increasing phylogenetic distance between species (Figure 3B). Relatively few regions were bound by any factor in all four species (Figure 3C). Overall both quantitative and qualitative comparisons led us to conclude that transcription factor binding has diverged considerably within the *Drosophila* genus.

## Differences of TF-binding motif content partly explain binding divergence

We next compared sequence divergence to binding divergence. We first mapped binding divergence along the six branches of the Drosophila tree (Figure 1), by modeling binding evolution according to a Brownian motion model [29–31]. This model is commonly used in evolutionary studies involving continuous quantitative traits [32–34] and is based on the hypothesis that a continuous trait (here TF binding) evolves neutrally and follows a Brownian motion from the ancestor to the daughter leaves. This model improves upon pairwise comparisons by taking into account the inherent phylogenetic inertia in our dataset (it corrects for the non-independence of traits due to phylogeny, [30]). We imputed binding values at the three internal nodes on the tree (Figure 1), including the root, and computed changes in binding along each of the six branches of the tree as the difference of binding values between parent and daughter nodes. We then compared these estimates of TF binding divergence to sequence divergence, modeled by the total number of nucleotide substitutions in the sequence of the corresponding bound regions along the tree. We only found a small correlation between binding divergence and overall sequence divergence (Figures 4A and S5).

We asked to what extent changes in the motif content of bound regions might account for the observed binding divergence. We first verified whether the TFs exhibited similar DNA binding properties in the different species (Figures 4B–C, S6 and S7), so that we could use the same motif (the known *D. melanogaster* motif, as established by [22]) to predict TF binding in all bound regions. The different orthologous TFs showed similar DNA affinities, although subtle differences in DNA binding properties could not be ruled out (Figure S6). Binding intensity in the sets of bound regions was then predicted, based solely on motif content and using our previously published thermodynamic model of protein-DNA interactions [35]. Binding predicted from motif content was imputed at the three internal nodes of the tree using the four species-specific binding predictions and a Brownian motion model. We found that changes in predicted binding, based on motif content, were correlated with TF binding changes branch-wise, both quantitatively (Pearson correlation from 0.16 to 0.24, p-value$<2.10^{-16}$, Figures 4D and S8A–D) and qualitatively (Figures 4E and S8E–H). Overall, these results suggest that changes in binding of BCD, GT, HB and KR are caused in part by changes in the distribution of TF-specific binding sites across the genome.

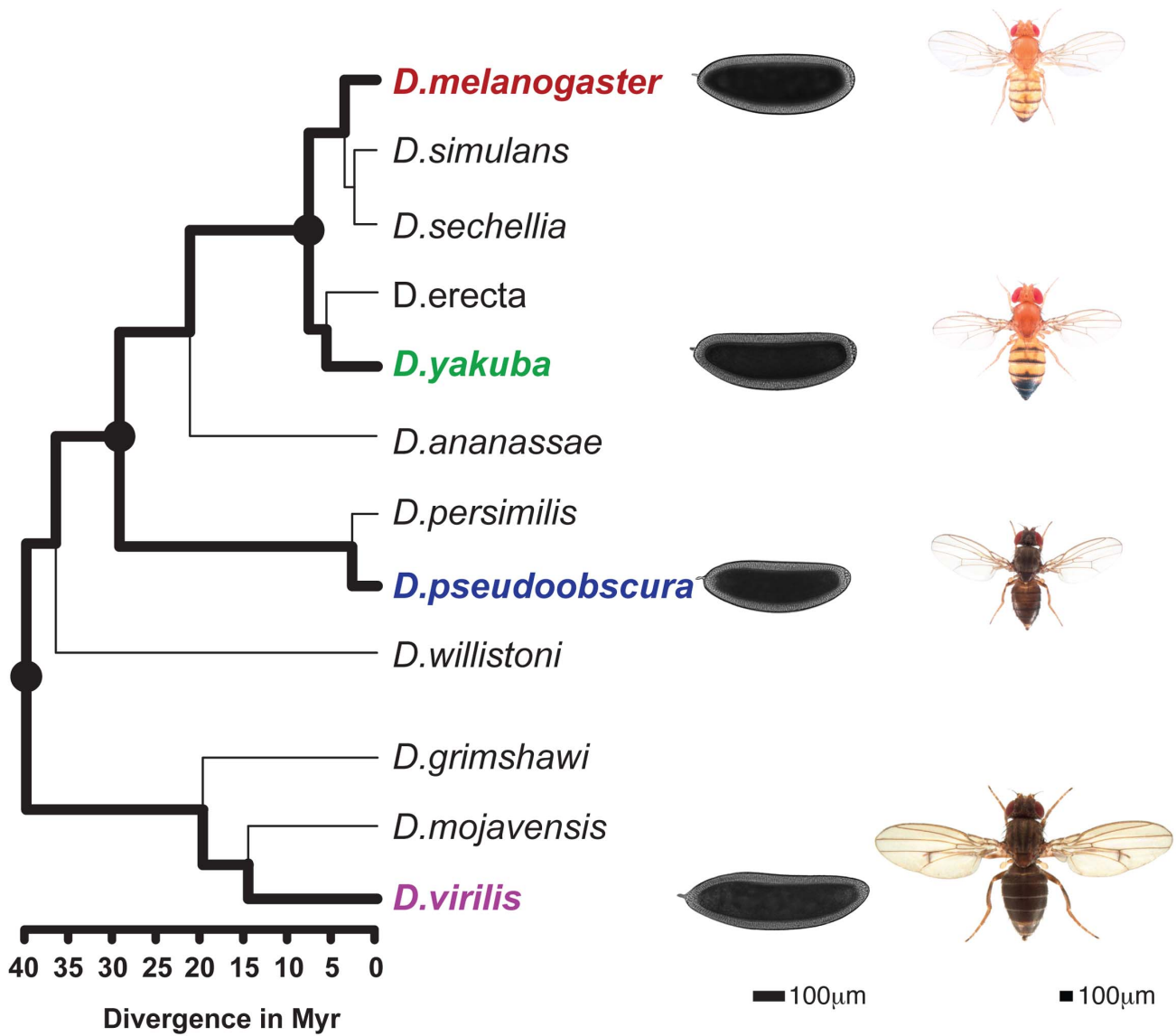

**Figure 1. Phylogenetic tree of the *Drosophila* genus.** The 4 species studied here (*D.melanogaster, D.yakuba, D.pseudoobscura, D.virilis*) are highlighted and illustrated with a picture of an adult as well a blastoderm embryo. Scales for adults (Nicolas Gompel, Flybase) and embryos are indicated. The three internal nodes of the (*D.melanogaster, D.yakuba, D.pseudoobscura, D.virilis*) tree are highlighted. Divergence times are indicated under the tree [20].
doi:10.1371/journal.pgen.1003748.g001

## Binding divergence is correlated between factors and associated with the transcription factor Zelda

We previously observed that the occupancy of the four factors (and others) is highly correlated in *D. melanogaster* [22,36], and *D. yakuba* [13]. We analyzed correlations amongst the binding of all four factors in each species independently or in all species using a principal components analysis of binding for all factors considered together. The first principal component (PC1), that explains the most variation in the data, affected all TFs in the same way in all species as well as in the dataset comprised of the binding data from all species (Figure 5B and S12). In other words the same correlation is present in each species, including in *D. pseudoobscura* and *D. virilis*, as well as in the combined dataset from the four species. In *D. melanogaster*, this axis is highly correlated with the DNA-binding levels of the protein Zelda (Zld) [37] (Figures 5C and S13), in agreement with previous studies showing that the correlation among the binding of different factors is driven by the early binding and chromatin-shaping activities of Zld [1,2,22,25]. Zld is present in the genome of all *Drosophila* species, its CAGGTAG binding site is enriched in regions bound by A-P factors in all species (Figure 5A), and the distribution of CAGGTAG motifs in bound regions predicts PC1 in all four species (Figure S13B). Overall these different elements suggest that Zld effect on chromatin and transcription factor binding is conserved among *Drosophila* species.

In addition to correlations in binding of different factors within a single species, we had previously found that changes in the binding of these AP factors were correlated between *D. melanogaster* and the closely related *D. yakuba*, and that this correlated divergence was driven by changes in Zld binding sites [1–6,13]. We repeated this analysis on the four species dataset and found that these relationships extend to the entire tree

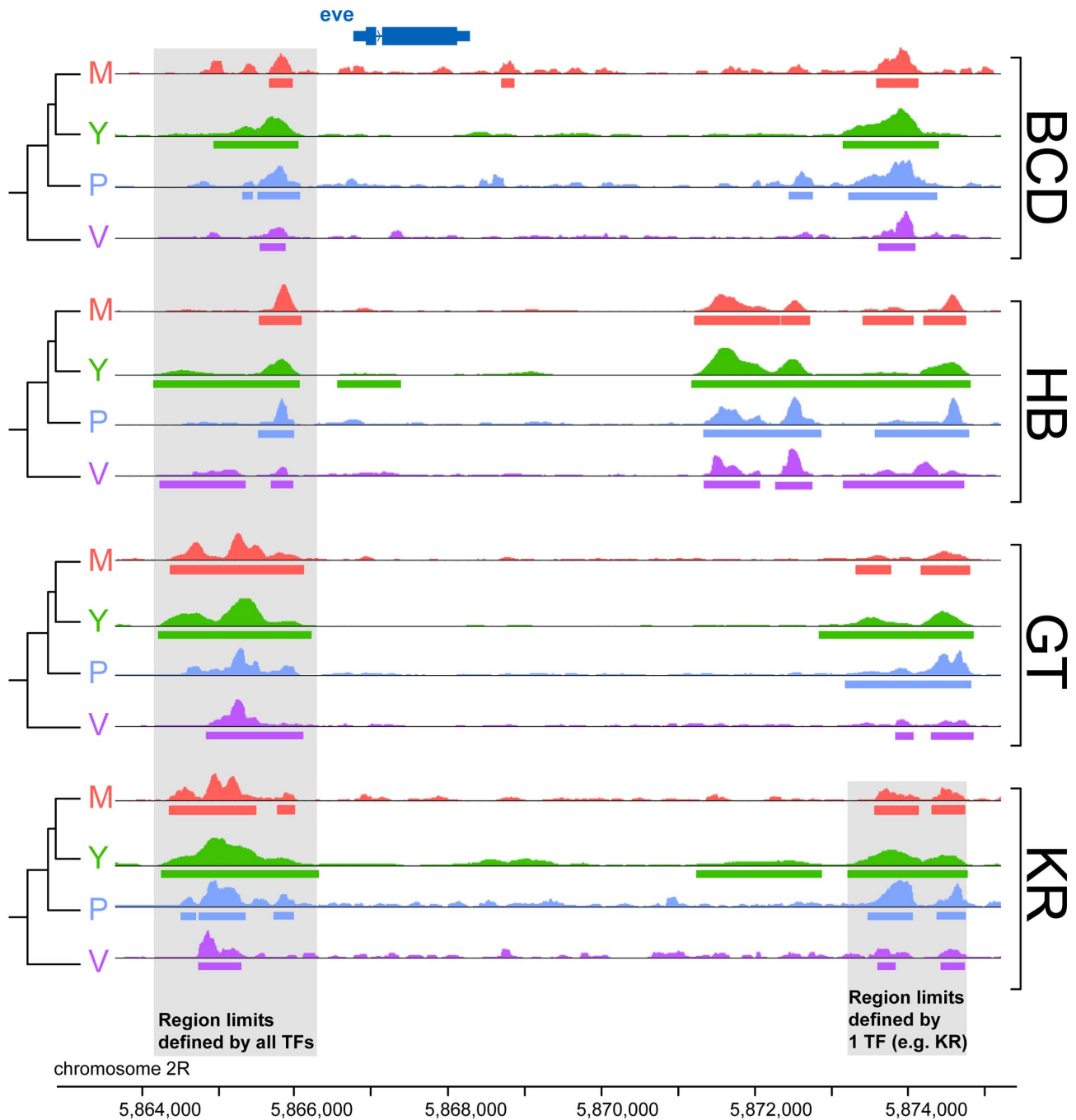

**Figure 2. Comparison of binding profiles of BCD, GT, HB and KR at the _even-skipped_ locus in the four species _D.melanogaster,_ _D.yakuba, D.pseudoobscura_ and _D.virilis._** An illustration of the two types of comparisons made in this study are highlighted in grey: trans-species comparison for each single TF (right) or trans-TFs comparisons (left). For simplicity, the species names were shortened using their initial: _D.melanogaster_ (M), _D.yakuba_ (Y), _D.pseudoobscura_ (P) and _D.virilis_ (V).
doi:10.1371/journal.pgen.1003748.g002

(Figure 5D–E). Although the effect is weaker, it remains significant.

### Regions likely to be functional enhancers are more likely to be conserved

We next examined the relationship between peak conservation and three additional factors that are all characteristics of known functional regulatory regions: peak height, location of peak relative to genes, and binding of other factors to the same region [7–14,22,36,38]. All three were correlated with peak conservation, as measured by the number of species in which binding was detected (Figures 6A, 6C, 6D), or the variance of binding estimated from the Brownian motion model (Figure 6B). High peaks, located outside coding regions and clustered among several TFs were much better conserved than small, isolated peaks located within coding regions. Of note, these covariates were not independent of each other – higher peaks are more likely to be clustered with peaks for other factors in non-coding regions. This result is independent of the thresholds used for identifying sets of bound regions (Figures S10 and S11). In addition, we found that binding in a set of active regions that function as A-P enhancers was significantly better conserved than in the rest of the genome (Figure S9). These "A-P" regions were composed of known A-P enhancers as well as sequences driving expression along the A-P axis (RedFly database [39]) and some of the sequences highly bound by many early embryonic factors [36] (subsequently dubbed Highly Occupied Target - or HOT - regions by [40]), which function as A-P enhancers in blastoderm embryos [38,41].

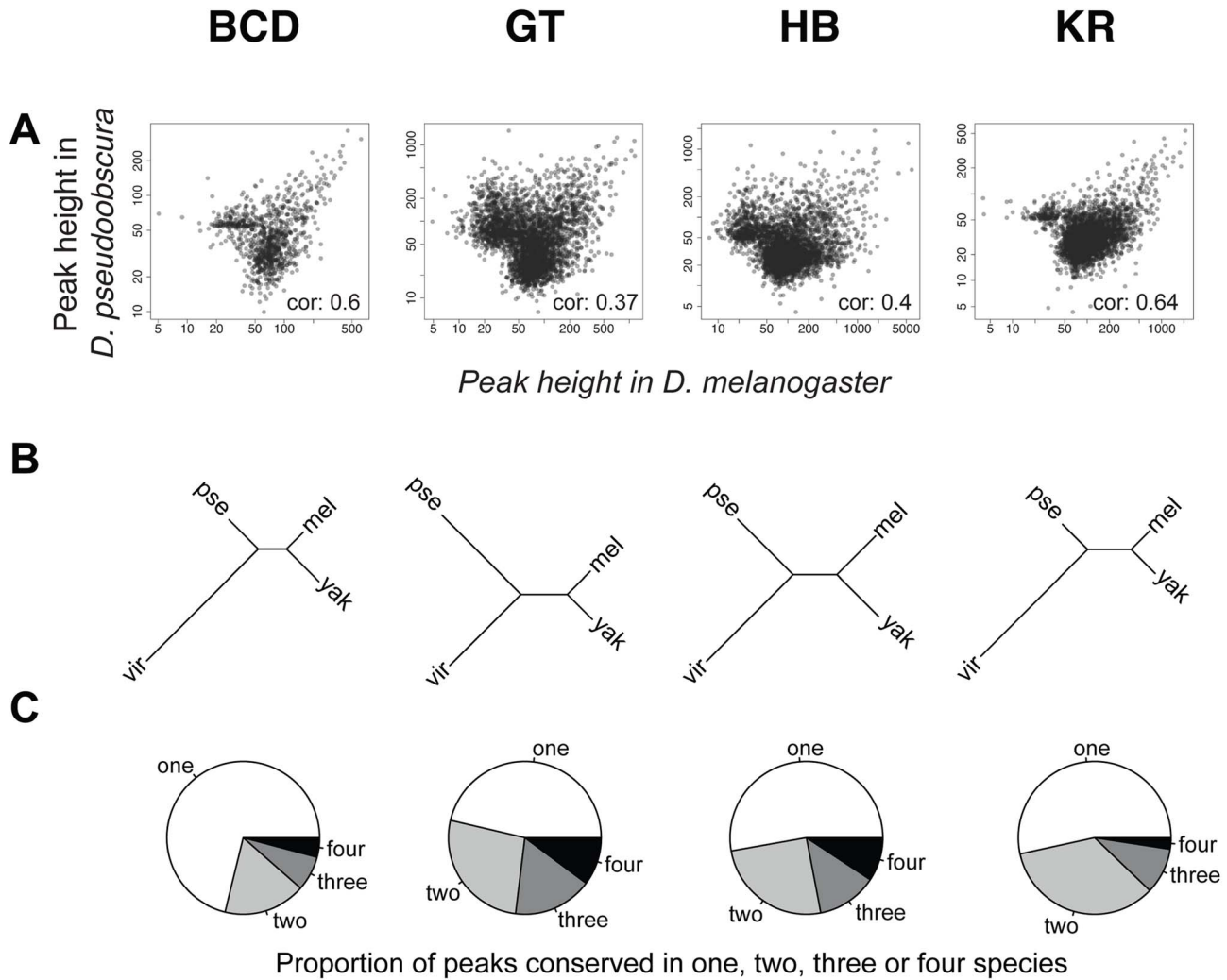

**Figure 3. Comparison of BCD, GT, HB and KR binding in *D.melanogaster, D.yakuba, D.pseudoobscura* and *D.virilis*. A**. Pair-wise comparisons of BCD, GT, HB or KR binding between *D. melanogaster* and *D. pseudoobscura*. Spearman's correlation coefficients are indicated. All correlations were highly significant (p-values$<2.10^{-16}$). See Figure S4 for all pair-wise comparisons. **B**. Neighbor-joining trees based on pairwise distance matrices of TF occupancy at bound loci. **C**. Proportion of the number of species for which TF was detected per cluster, from a species-specific peak ("one"), to a peak conserved in all 4 species ("four"). For simplicity, the species names were shortened using their first three letters: *D.melanogaster* (mel), *D.yakuba* (yak), *D.pseudoobscura* (pse) and *D.virilis* (vir).
doi:10.1371/journal.pgen.1003748.g003

## Higher proportion of binding variation explained within species than between species

We applied a simple multiple linear regression model to all of the parameters described above (TF-specific and Zelda motif enrichments, proximity to genes, number of other TFs binding the same locus) and found that, while these factors explain about 29–36% of the variance in binding within each species (Pearson correlations ~0.6, p-value$<2.10^{-16}$, Figure S14 A–D), they only explain 3–7% of the variance in TF binding divergence between species (7–21% if phylogenetic inertia is not taken into account, Figure S14 E–L).

## mRNA levels are better conserved than TF binding

To investigate how the level of divergence of transcription factor binding affects gene expression, we sequenced mRNA from embryos from each species harvested at the end of cellularization using high-throughput mRNA sequencing (mRNA-seq). The reference annotations (Flybase) in non-*D. melanogaster* species are of lower quality (e.g. the annotated *D. yakuba, D. pseudoobscura* and *D. virilis* transcriptomes cover ~22.7, 23.5 and 21.8 millions nucleotides, respectively, compared to 31 millions for *D. melanogaster* when similar sizes are expected). As this quality

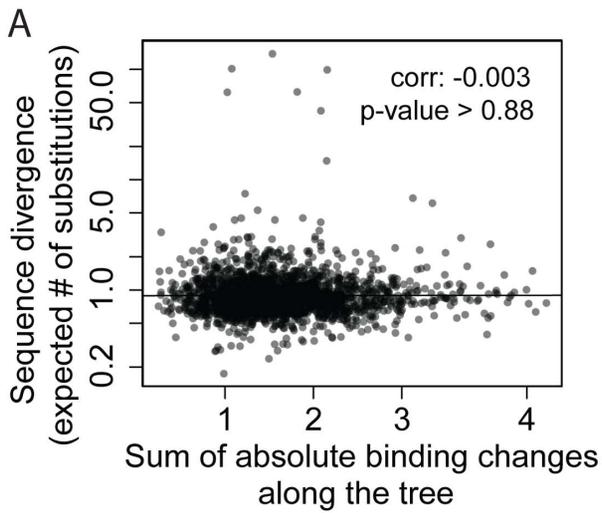

A

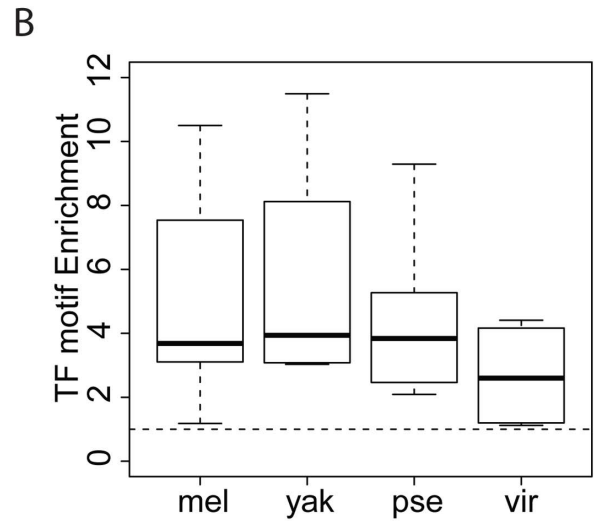

B

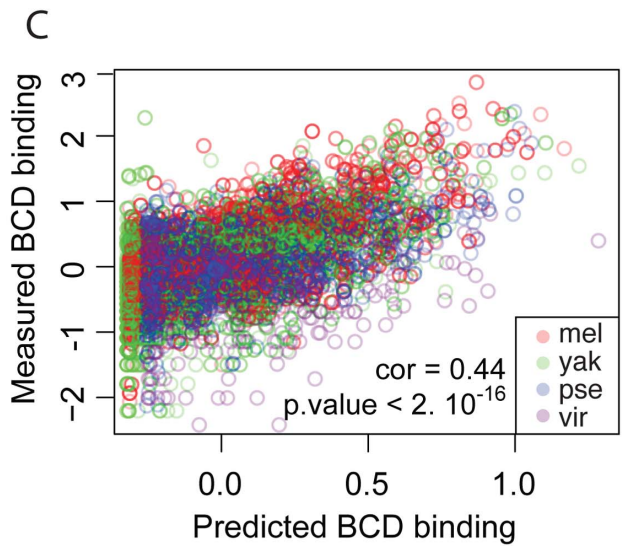

C

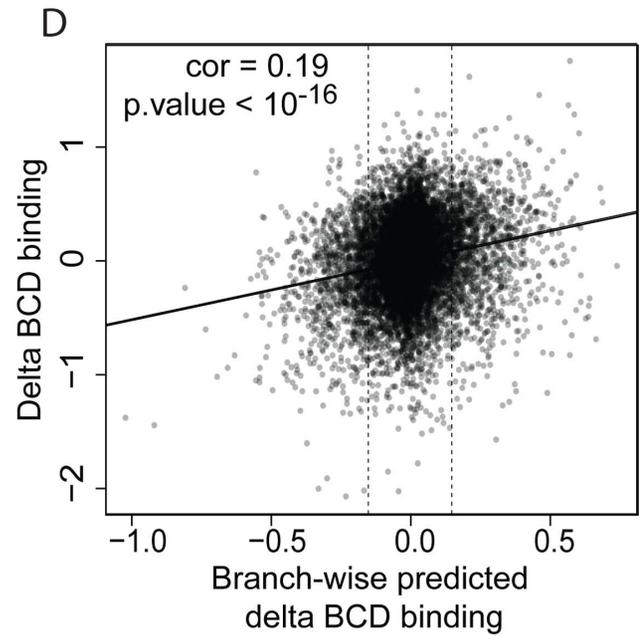

D

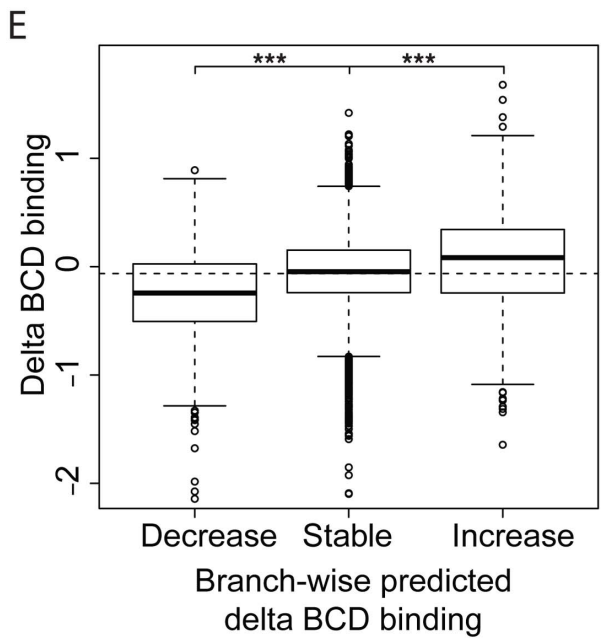

E

**Figure 4. TF-specific motif turnover drives TF binding divergence. A**. Comparison of quantitative variation of BCD binding divergence *vs.* underlying sequence divergence. Binding divergence was measured by the variance of BCD binding along the *Drosophila* tree according a Brownian motion model. Sequence divergence was measured as the total length of a PhyML phylogenetic tree based on the sequence alignment underlying bound regions. **B**. Sequences under bound regions are enriched for TF-specific motifs in all species. TF-specific enrichment was calculated for each of the 28 ChIPs. The plot summarizes motif enrichment in any of the 28 ChIPs distributed between 12, 8 and 4 ChIPs in *D. melanogaster*, *D. yakuba*, *D. pseudoobscura* and *D. virilis*. **C**. Enrichment of BCD motifs in bound regions is quantitatively highly predictive of BCD binding. BCD binding prediction was solely based on underlying BCD motif content (TAATCC) [35] **D**. Comparison along the tree of branch-wise BCD binding divergence and predicted BCD binding divergence, **E**. Values of BCD binding divergence (same as **D**) were partitioned into three categories, depending on predicted changes of BCD binding along a branch: predicted increase, decrease or limited change in binding (thresholds indicated by vertical lines in **D**.). ***: Wilcoxon test p-value<0.001. Similar plots for GT, HB and KR can be found in Figures S7 and S8.
doi:10.1371/journal.pgen.1003748.g004

difference may induce biases in our analyses, we first refined the reference annotations for the non-*melanogaster* species. For this, we used the RABT program from cufflinks (option –g, [42]) and mRNA-seq data from pools of ~50 embryos spanning the same stages used in the ChIP experiments. This method builds on an existing annotation and mRNA-seq data to discover novel transcripts. The number of genomic nucleotides covered by the annotations, increased by 17%, 19% and 24% for D. *yakuba*, *D. pseudoobscura* and *D. virilis*, respectively, and are more similar to *D. melanogaster* values (Table S4). We used these new annotations plus the *D. melanogaster* reference annotation for subsequent experiments.

We conducted mRNA-seq experiments on several individuals from each species and restricted our analyses to the 8,555 protein-coding genes for which we could identify orthologs present exactly once in all four species using our genome alignment. As for ChIP experiments, log-transformed values were normalized so that median was null before further comparisons of levels of gene expression. We found that mRNA levels are highly correlated between species (Figure 7A), with the correlation decaying with phylogenetic distance (Figure 7B). In addition, overall divergence of mRNA levels was significantly lower than divergence of any TF's binding, as measured by the variance of normalized data along the *Drosophila* tree (Figure 7C).

The magnitude of this difference should be interpreted with some caution, however, as ChIP is intrinsically noisier than mRNA-seq (correlations of ChIP replicates ranging from 0.45 to 0.91 with a median at 0.81 *vs* correlation of mRNA-seq replicates ranging from 0.95 to 0.98 with a median at 0.97) and involves using reagents, such as anti-TF antisera, that could have subtle differences in activity among divergent fruit fly species, which may introduce some additional variance relative to actual differences in binding.

## TF binding divergence is poorly correlated with divergence of mRNA levels

To compare variation in transcription factor binding to variation in gene expression on a gene-by-gene basis, we matched each gene to the closest region bound by the highest number of different TFs (one of the 10,137 merged regions), recognizing that some of these associations were likely to be incorrect. We focused on the 4,846 genes that were annotated in all four species and expressed in at least one of them. Of these, 3,024 could be associated with nearby binding of at least one transcription factor in one species. We partitioned genes depending on whether transcripts in *D. melanogaster* blastoderm embryos were deposited by the mother into the egg, or were a product of zygotic transcription, as defined by [43]: 2,056 genes were categorized as "maternal", 388 as "zygotic" and 394 as "both maternal and zygotic" as they are both deposited by the mother and transcribed by the zygote; the 186 remaining genes were not categorized.

We analyzed for each gene the relationship between mRNA levels and nearby TF binding. TF binding, and especially strong and clustered binding, was preferentially localized near zygotic genes (Figure 7D). We found a positive correlation between measured mRNA levels of zygotic genes and mRNA levels predicted solely by a multiple linear regression of associated nearby TF binding (Pearson correlations within each species ranged from 0.42 to 0.5. Using combined data from all species, correlation is 0.46, p-values<$10^{-16}$, Figure 7E). Within a species, 17% to 23% of the variance in mRNA levels could be attributed, by multiple linear regression, to variance in levels of associated TF binding. Association between mRNA levels and nearby TF binding for maternally deposited genes was comparatively much weaker and intermediate for "maternal/zygotic" genes (Figure S15). This result is coherent with the known role of TFs as transcriptional regulators of zygotic gene expression but not of maternal gene expression.

Despite this relatively good relationship between TF binding and gene expression within a species, we found a weak relationship between variation in TF binding and gene expression along the tree (Figure 7F). Using multiple linear regression we found that trans-species variation of mRNA levels is positively correlated with trans-species variation in associated TF binding (Pearson correlation ~0.16, p-value<$10^{-16}$), but only ~2% of variance in mRNA levels could be attributed to variation in TF binding for zygotic genes (3% if phylogenetic inertia is not taken into account, Figure S16). In contrast, divergence of mRNA levels for maternal genes was not correlated with changes of associated TF binding, suggesting that the effect seen for zygotic genes, although weak, is of biological significance. Accordingly, binding of A-P factors near zygotic genes is better conserved than near maternal genes (Figure S17). We finally compared the expression pattern for three genes displaying exhaustive changes in associated TF binding. We selected zygotic genes displaying a known A-P pattern and high levels of associated TF binding in *D. melanogaster* but comparatively much lower levels of binding in at least another species. For two out of three cases, we found changes of expression along the A-P axis correlated with binding changes (Figures S18, S19, S20).

## Discussion

This study addresses both the genomic causes and the consequences on gene expression of TF binding divergence. We found that the genomic binding of A-P factors has changed extensively along the *Drosophila* tree with little effect on downstream gene expression. We identified two potential genetic sources of binding divergence. First, we found that alterations of BCD, HB, GT and KR-specific motifs drive a portion of the binding divergence. This was previously found for other TFs [7,10,12,13,44,45], though our study is among the first to quantify this effect. We found that the A-P factors are undergoing correlated evolution among themselves (Figures 5D and 6D) and

**A**

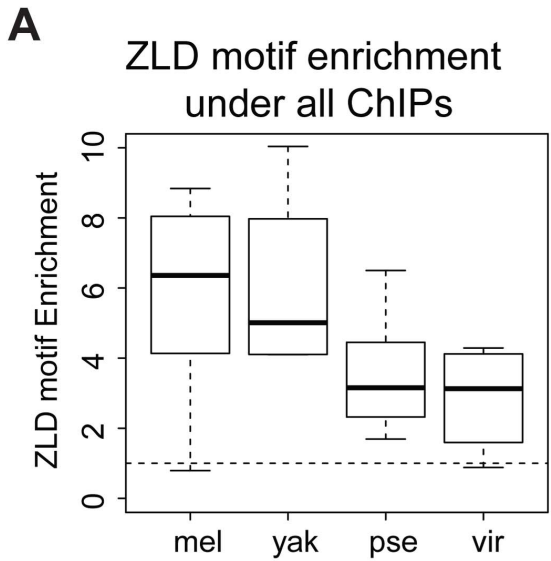

**B**

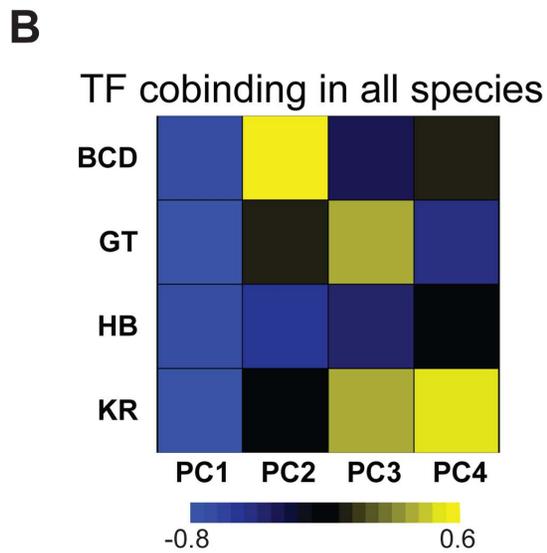

**C**

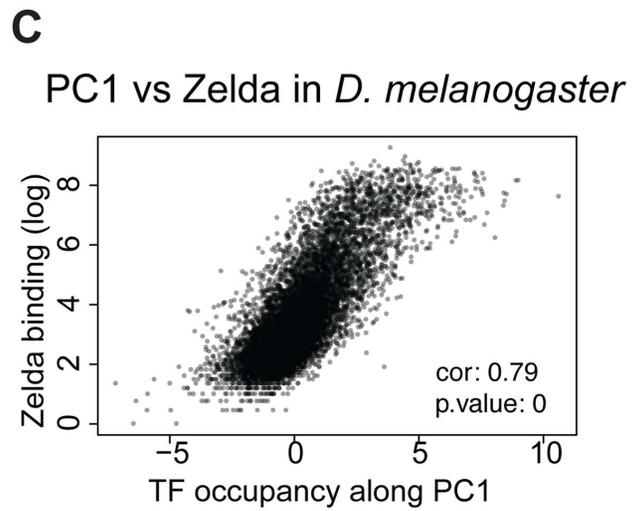

**D**

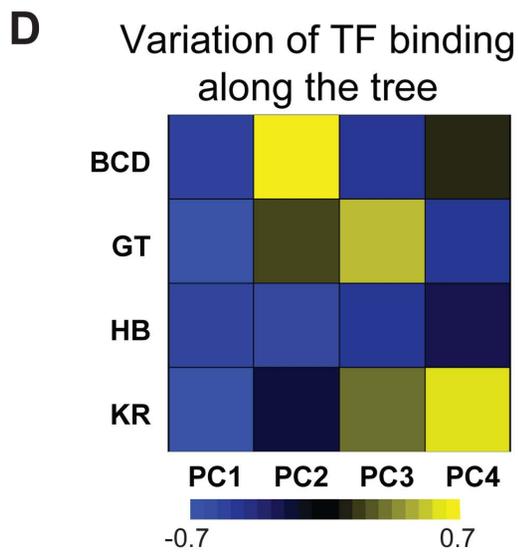

**E**

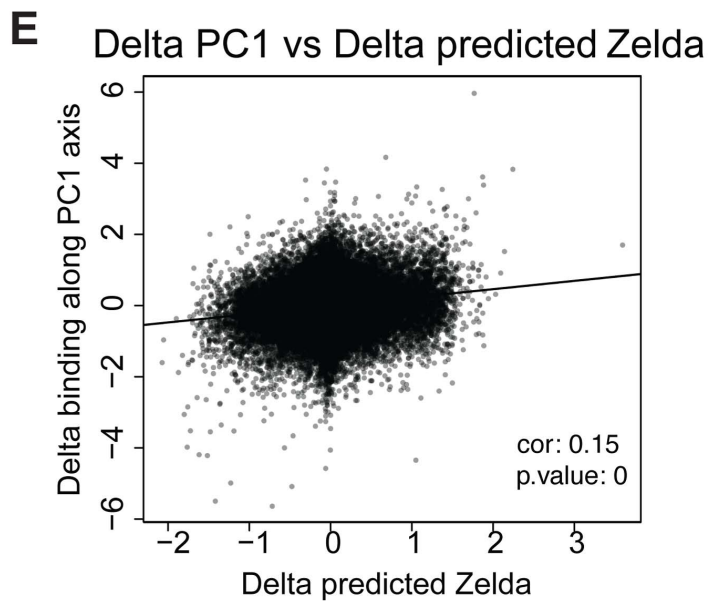

**Figure 5. Zelda divergence may drive TF binding divergence. A**. Zelda CAGGTAG motif is enriched in all ChIPs in all species. The enrichment variability between species is mostly due to differences in ChIPs qualities. **B/D** Principal Component Analysis (PCA) of binding of all factors. (**B**) PCA of the binding strength in the 4 species together and (**D**) PCA on the change in binding strength along each branch of the tree across all peaks. Each row represents a factor, and each column is a principal component of the relevant data. The color represents the sign (yellow positive, blue negative) and magnitude (color intensity) of each value in each principal component. In each case the sign of the first principal component is the same for all four factors, indicating that the dominant driver of both interspecies divergence and quantitative variation within single species is a coordinated change in binding strength of all factors. This effect explained 40% of the variation between species, and 58% of the variation within species. Species-specific PCAs are shown in Figure S12 **C**. Zelda occupancy (from [25]) and PC1 coordinates are highly correlated (Spearman correlation ~0.79, p-value$<$10$^{-16}$). **E**. Changes in PC1 along the branches of the tree correlate with changes in Zelda binding predicted solely by the enrichment of the Zelda motif. *** Wilcoxon test p-value$<$0.001. See Figure S13 for comparison of predicted Zld binding and Zld binding.
doi:10.1371/journal.pgen.1003748.g005

potentially with the protein Zelda (Figure 5E), pointing to the importance of *cis*-acting sites beyond target sites for the factors themselves in binding divergence.

Given the high degree of morphological stability amongst these four species, we predicted and observed highly conserved embryonic gene expression – for both maternal and zygotic genes. Yet, conservation of gene expression was not coupled with conservation of genome-wide transcription factor binding.

One relatively simple explanation is that our measurements of transcription factor binding are more subject to systematic biases and noise, since ChIP experiments are generally less robust than mRNA-seq and involve species-specific reagents. The inherent experimental noise of ChIP experiments will thus lead to overestimating binding divergence, although it is difficult to quantify its contribution. This said, we believe our conclusions are robust to ChIP noise, if only because the same conclusions were drawn from ChIPs of different qualities (GT and HB ChIPs were the best datasets whereas BCD and KR ChIPs showed lower enrichment and reproducibility, Figure S3). It is also possible that, even though we carefully optimized embryo collections to account for the different development times of the species, we may not have perfectly matched stage sampling between species. However, the divergence of TF binding we observe is in line with most genome-wide comparisons of TF binding, including in *Drosophila* [7–10,12,14,46], although these studies also face the same limitations that ours does. Our recovery of motif signals for BCD, GT, HB and KR in their bound regions, and ZLD in all of them, argues against extreme levels of noise in the data. Overall some of the difference in divergence (Figure 7C) undoubtedly originates from both the different nature and means of measurement of TF binding and gene expression.

Our interpretation of extensive binding divergence of A-P factors contrasts somewhat with a recent comparison of the binding of the dorsal-ventral patterning factor Twist [47]. He et al. found that 34% of the regions bound by Twist in *D. melanogaster* are also bound in five species ranging from the closely related *D. simulans* to the more distant *D. pseudoobscura*, leading the authors to claim that Twist binding was highly conserved. In comparison, we find that 28%, 38%, 23% and 15%, of BCD, GT, HB and KR regions bound in *D. melanogaster* are also bound in *D. yakuba* and *D. pseudoobscura*, which correspond to roughly similar levels of binding conservation. Thus we differ primarily in whether the finding that only one of three bound regions found in *D. melanogaster* is conserved constitutes a high degree of conservation. More interestingly, the Twist study confirmed our results that functional binding is more constrained than seemingly non functional binding since clustered Twist binding and binding near target genes were better conserved [47].

Perhaps the strongest argument that the overall low levels of conservation we observe is real is that we observe strong conservation for the subset of bound regions that are most likely to be functional. This suggests, unsurprisingly, that there is strong purifying selection to maintain binding in functional regions. But it also demonstrates that we can detect strong conservation when it is there, with the corollary that the regions we observe to be divergent are likely to really be divergent. That two thirds of the regions bound by any of the four TFs we examined are poorly conserved and thus are probably under weak or no purifying selection supports the emerging view that a large fraction of measureable biochemical events are not functional (in contrast to claims made by ENCODE [48]). We note, however, that strong selection to maintain binding does not require conservation of individual binding sites [45,49–53] There is also evidence that even within highly conserved regulatory networks, differences in embryo size and other factors necessitate constant tuning of enhancer activities, with corresponding adjustments in the strength and organization of TF binding sites [54]. Unfortunately, the ChIP data we report here lacks sufficient resolution to see the turnover of individual binding events. New techniques like ChIP-exo [55] enable much higher resolution moving forwards, and the growing repertoire of sequenced and experimentally tractable *Drosophila* species should allow us to study turnover at both the sequence and binding level with much more precision.

## Methods

### Antibodies used for ChIP

We used rabbit polyclonal antibodies raised against the *D.melanogaster* versions of the key A-P regulators Bicoid (BCD), Hunchback (HB), Giant (GT) and Krüppel (KR) that had been produced in a previous study [22]. They were affinity purified either against the *D.melanogaster* version of the proteins (recognizing the largest set of epitopes), or against the most distantly related version from *D.virilis* (recognizing the most highly conserved epitopes). The different proteins are high ly conserved throughout the *Drosophila* gender (69%, 87%, 76% and 70% aminoacid identity between *D.melanogaster* and *D.virilis* BCD, HB, GT and KR epitopes, respectively), allowing an excellent cross-reactivity of the serum against proteins from all species.

### Embryo collections for ChIP

We collected embryos spanning early cellularization process (end of stage 4 to mid stage 5), during which the regulatory events that initiate segmentation along the A-P axis take place. Because the developmental speeds and optimal growth temperatures vary between species, the collection conditions were optimized for each species (Table S1). Embryos from all species were processed either for mRNA-seq, as described below, or for ChIP-seq (except *D.yakuba*), according to the protocol described below. In addition, we collected single embryos for each species, at the end of cellularization, just prior to gastrulation, based on morphological criteria, homogenized the embryos in TRIzol (Life Technologies) and processed the samples for DNA and total RNA extraction, as described below.

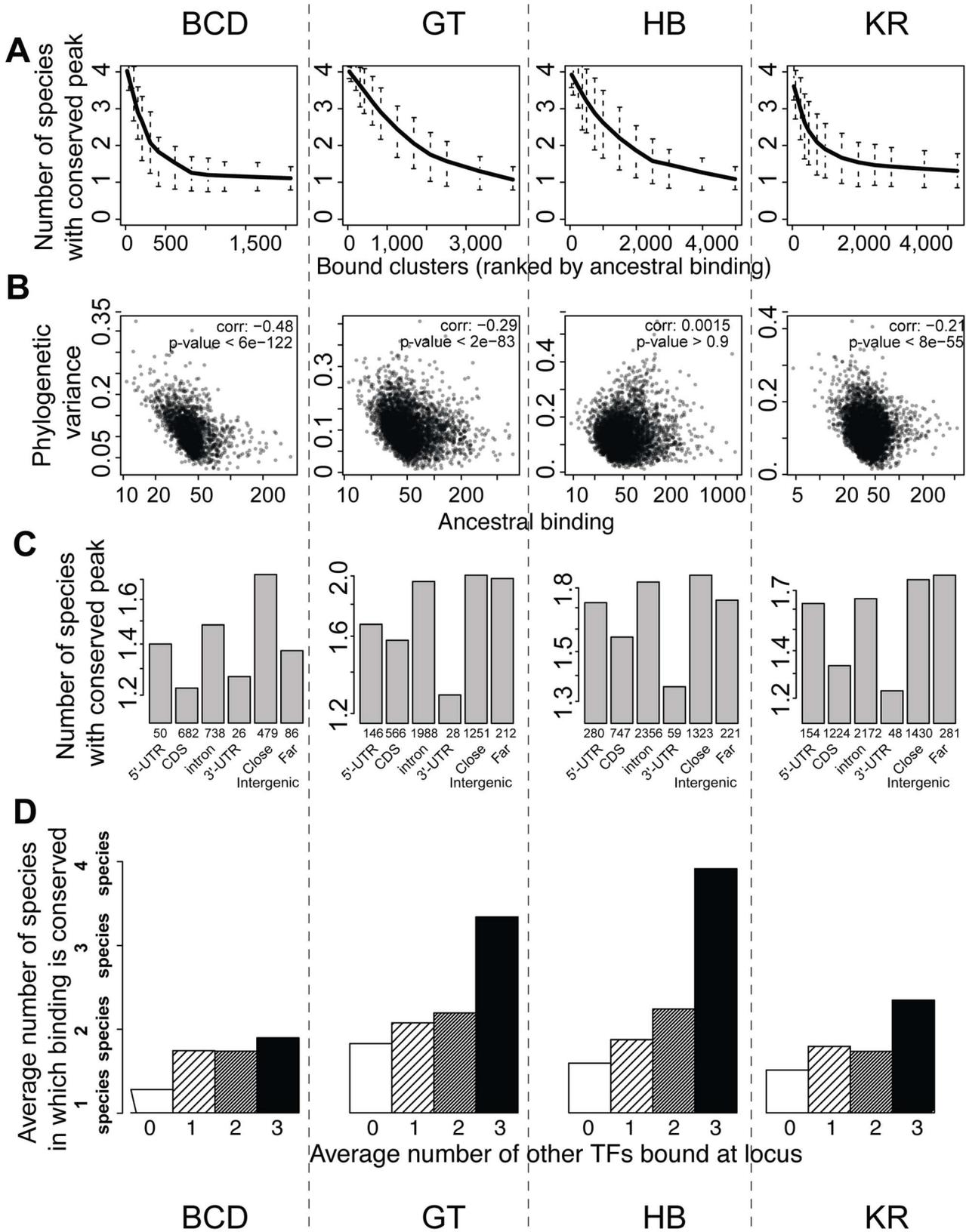

**Figure 6. BCD, GT, HB and KR binding events are differentially conserved, and binding predicted to be functional is better conserved. A**. Comparison of qualitative conservation of TF binding in the different species. A conservation score, corresponding to the average number of species in which binding was detected (1–4), was calculated for each set of orthologous regions and ranked according to ancestral mean, as estimated using a Brownian motion model. **B**. Comparison of binding intensities, as represented by ancestral mean binding, and trans-species

binding variance in a Brownian motion model of TF binding evolution. **C.** Mean binding conservation score (1–4 species) depending on peak location in *D. melanogaster*. **D.** Mean binding conservation score depending on the number of other A-P factors binding the same locus. To correct as much as possible for TF binding differences linked to different wiring sizes, clusters were binned into 10 bins, depending on the estimated ancestral values, and the conservation was estimated independently in each bin. The average conservation is displayed. Similar plots were made using different threshold for calling sets of bound orthologous regions (Figures S10 and S11).
doi:10.1371/journal.pgen.1003748.g006

## In Vivo formaldehyde cross-linking of *D.melanogaster*, *D.pseudoobscura* and *D.virilis* embryos, followed by chromatin immunoprecipitation

Embryos were collected as described above and fixed with formaldehyde. The chromatin was isolated through CsCl gradient ultracentrifugation as previously described [22].

The chromatin used for immunoprecipitation was fragmented through sonication using a Bioruptor to an average fragment size of 180 bp. After sonication ChIP was carried out using affinity purified rabbit polyclonal antibodies directed against large parts of BCD, HB, GT or KR, and prepared as described above: antibodies affinity-purified against the *D.melanogaster* epitope [22] or against the *D.virilis* orthologous parts of the *D.melanogaster* epitopes (for BCD, HB and KR). Both sets of antibodies essentially give similar results (GT and HB giving the most reproducible results), although the *D.virilis*-specific antibodies were less efficient, as expected since they recognize the smaller set of conserved epitopes compared to the other ones (lower signal over noise ratio) (see Figure S3 for qualitative and quantitative comparison of replicates). Some samples from *D.melanogaster*, *D.pseudoobscura* and *D.virilis* (labeled with an asterisk in Figure S1) were pooled prior to immunoprecipitation, in order to minimize variation due to sample handling and protocol adherence.

The DNA libraries for sequencing were prepared from the ChIP reaction and from Input DNA following the Illumina protocol for preparing samples for ChIP sequencing of DNA. All library amplifications were carried out by 15 cycles of PCR. After the amplification step, we size-selected DNA fragments of 190–290 bp. Library quality, fragment size, and concentration were measured as described for mRNA-seq. Libraries were sequenced on a GAIIx or HiSeq 2000 Illumina sequencer.

### mRNA-seq library preparation

Individual embryos from late stage 5 were chosen for RNA extraction based on morphology (having completed cellularization), which allows for sampling of homologous stages across species. Embryo sampling was performed as described in [43]. Libraries were then made from total RNA of 6 individuals of each species, for a total of 24 libraries, using the mRNA TruSeq kit from Illumina, following the manufacturer's instructions. Size distribution of the library fragments was checked on a Bioanalyzer (Agilent) using the high sensitivity kit, and library concentration was measured by QPCR using Kapa Biosystems PCR kits for Illumina sequencing libraries, according to the manufacturer's instruction. Libraries were sequenced in 2 lanes (12 libraries per lane) using an Illumina HiSeq sequencer. mRNA levels were later averaged over the 6 individuals from the same species.

In parallel, pools of embryos spanning the end of stage 4 to mid stage 5 were collected (see Table S1). After RNA extraction, samples were processed for library preparation, as described above and paired-end libraries were sequenced on an Illumina GaIIx. These samples were only used for modifications of annotations (see below and Table S4).

## Genome versions and mapping sequenced tags to genomes

We used the Apr. 2006 assembly (Flybase Release 5) of the D. melanogaster genome, the February 2006 assembly (Flybase release 2) of the D.pseudoobscura genome and the February 2006 assembly (Flybase release 1) of the *D.virilis* genome.

The *D.yakuba* data as well as some of the *D.melanogaster* dataset were previously published [13].

We trimmed all sequenced tags so that their average quality was above 30 and mapped the tags to the genomes using Bowtie v0.12.7 [23] with command-line options '-v 1 -m 1' for small reads (length below 35 bp) and '-v 1 –m 3 for long reads (length above 70 bp), thereby keeping only tags that mapped uniquely to the genome with at most one or three mismatch. For ChIP-seq experiments using chromatin pooled from several species, we used only reads that could be unambiguously assigned to a single species. To this end, reads were separately mapped to each genome sequence and the reads that mapped to several genomes were discarded. Reads that mapped better to a genome sequence in particular (with at least 2 mismatch differences), were recovered in a second time. More than 99% of the reads could unambiguously mapped to a genome sequence. Overall pooling had very limited effect on downstream read mapping because these species are distantly related, and only a handful of regions were affected by pooling (Figure S2).

### Peak calling

ChIP data was parsed independently for each experiment using two separate peak callers. First, we used MACS (version 1.4) [26], with the following parameter "-g dm –off-auto –nomodel – pvalue = 1e-2" and "–shiftsize = 110 –mfold = 10,10000 –slocal = 2000 –llocal = 20000" or "–shiftsize = 60 –mfold = 4,10000" depending on the length distribution for DNA fragment sizes prior to sequencing. We also called peaks using Grizzly Peak fitting program [24,25] with estimated DNA fragment length of 150 or 250.

We then intersected the two sets of peaks, and filtered out all peaks not supported by both methods. To account for low complexity peaks and possible PCR artifacts, we further removed peaks with negative correlation ($<-0.1$) among the Forward and Reverse reads, peaks where 60% of the reads mapped to less than 1% of the positions, and peaks whose height was less than three times the height of Input reads in the same locus.

We took as an initial dataset the union of all bound regions in the different replicates. Roughly similar numbers of peaks were identified in each species (Table S3), except in the case of *D.pseudoobscura* that displayed less BCD, GT and KR peaks, potentially partly due to coverage differences on the Müller element E, fused to the X chromosome in the *Pseudoobscura* group [5,56].

### Whole-genome alignment and orthology comparisons

To produce the 4-species whole genome alignment, we followed the general guide-line described in [57]. We used a large-scale orthology mapping created by Mercator [27] with the option to identify syntenic regions of the genomes. Each region was then aligned with Pecan [28] with the default options.

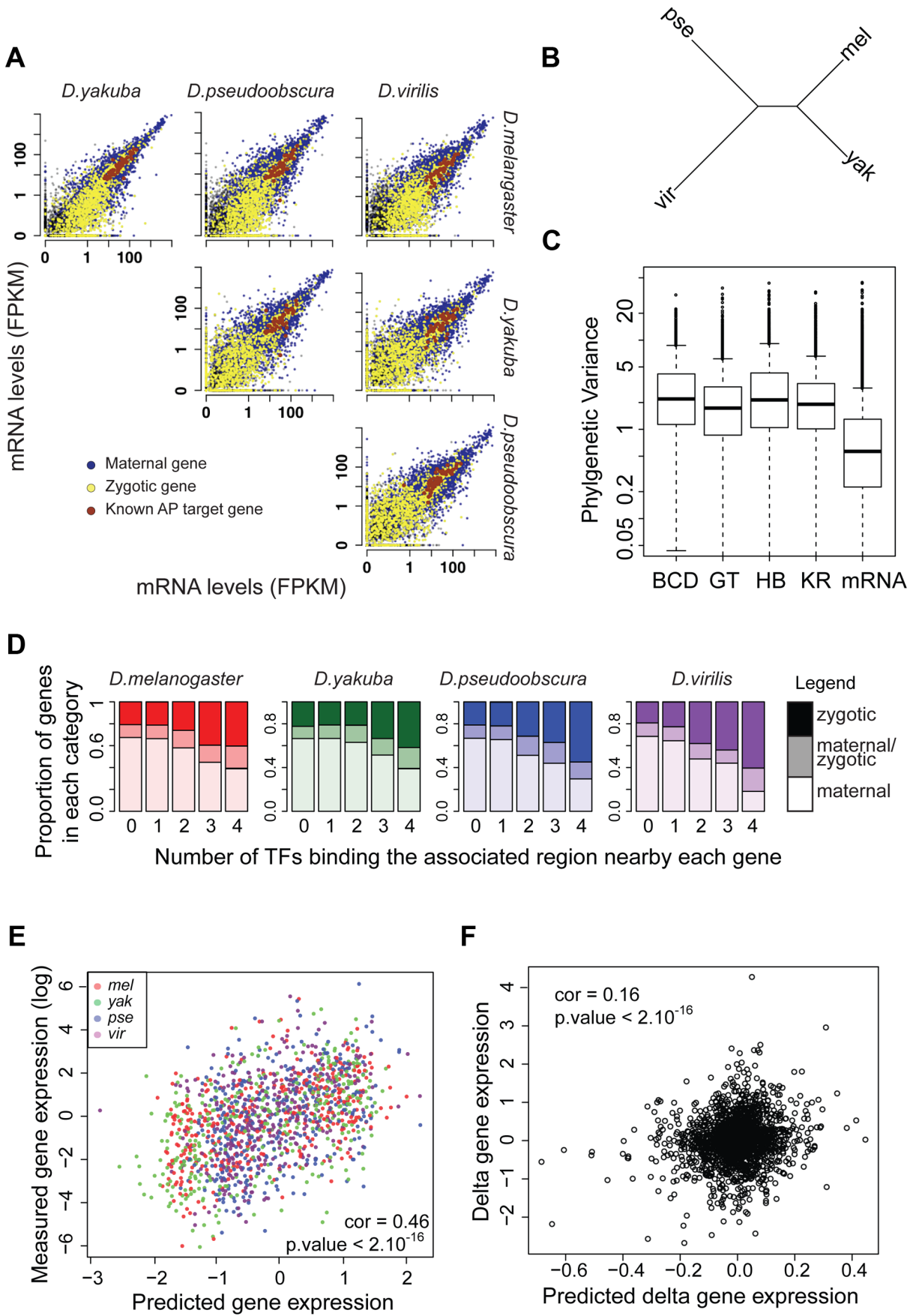

**Figure 7. mRNA levels are highly conserved despite high divergence of BCD, GT, HB and KR binding. A**. Pairwise comparison of mRNA levels in *D.melanogaster*, *D.yakuba*, *D.pseudoobscura* and *D.virilis* blastoderm embryos. **B**. Neighbor-joining tree based on pair-wise distance matrices of mRNA levels, based on Spearman's correlation coefficient. **C**. Phylogenetic variance of mRNA levels is significantly lower than variance of BCD, GT, HB or KR binding (Wilcoxon test p-value<$10^{-16}$). In order to compare variance, quantitative values were normalized by dividing each dataset by its standard deviation, on which parameters of the Brownian motion model were reestimated. **D**. Proportion of bound regions associated with maternal, zygotic or maternal-zygotic genes, depending on the number of TFs binding the region. Regions bound by many TFs tend to be localized near zygotic genes whereas isolated peaks tend to be localized near maternal genes. **E**. mRNA levels of zygotic genes are well predicted by associated TF binding in all species. mRNA levels were predicted from a multiple linear regression of associated nearby TF binding. **F**. Changes along each branch of the tree of mRNA levels for zygotic genes are modestly but significantly correlated with predicted changes based on quantitative changes of associated TF binding.
doi:10.1371/journal.pgen.1003748.g007

## Establishing the set of bound regions from a 4-way genome alignment

Bound regions were considered orthologous according to the following pair-wise rule: bound regions in two different species are considered orthologous if they share a non-null intersection on the alignment (see Figure 2). Regions were removed from the analysis based on quality check: (i) Clusters that displayed a genomic length variability above five fold between any two species were considered to have an unreliable alignment, (ii) clusters for which less than 80% of genomic length from any species could be unambiguously mapped on the sequence were considered to have too ambiguous occupancy values and (iii) sets of orthologous peaks comprised of a small peak (normalized value below 0.7) in only one of the seven replicates were considered to be too uncertain.

For each region, TF occupancy was obtained as the mean maximum occupancy (at the log-scale) between replicates in each species-specific ChIP. Datasets were then centered so that median was null.

We adopted a conservative approach when comparing binding between species: we used 2 different thresholds for calling a region "bound" and a peak "not conserved". We used a relatively high threshold for calling a set of orthologous bound regions (*ie* to look at the region) and we used a lower threshold to call peaks in this region so that a high peak and a low peak are more likely to be both called and thus considered conserved. This method allows us to minimize false-positive peak calling (that is likely to be divergent), and maximize conservation call. Our results were independent from the high threshold: we also repeated some of our analyses using different thresholds for calling sets (but keeping the same low threshold for determining binding conservation), with similar results (Figures S10 and S11).

## Inference of sequence divergence under bound regions

Alignments of the sequences corresponding to the sets of 2061, 4191, 4986 and 5309 orthologous regions bound by BCD, GT, HB and KR in at least one species were retrieved from the four species genome alignment. A phylogenetic tree was built per set using PhyML [58] with the parameters -m gtr -v e -t e -c 4 -a e -f m -d nt -o lr –quiet and the input topology (mel:0.119585, yak:0.13237, (pse:0.53, vir:1.0325):0.71552); obtained from the *Drosophila* genome project. Sequence divergence was measured as the total number of substitutions, and given by the total length of each output phylogenetic tree.

## Clustering of bound regions

The peaks called for every ChIP were further clustered, based on their mapping to the four-way alignment of the genomes. Overlapping peaks for the same TF over two or more species were merged.

## Motif enrichment analyses

To support the binding data, we analyzed the sequences of bound regions, and calculated the enrichment of each factor's motif. For every ChIP experiment (namely, every species/TF), we analyzed the bound regions using our previously published thermodynamic model of protein-DNA interactions [35]. For simplicity, we used fixed PWMs of BCD, GT, HB, KR and Zld from *D. melanogaster* for all four species [25,35] (BCD, GT, HB and KR motifs are displayed on top of Figure S6). Similar results and PWMs, were obtained when allowing for species-specific or experiment-specific optimization of PWM for each AP factor.

Average motif enrichments displayed on Figures 4B and 5A were measured on the top 250 peaks. Motif enrichment displayed in Figure S6 was measured as follows: for every bound region, we estimated the sequence-based probability of binding along every position. We then aligned these 500 top regions for each experiment based on the strongest site, and calculated the average predictions over surrounding windows of −500 to +500. As control, we repeated this analysis over the same loci, but used shuffled versions of the PWMs. Specifically, we first shuffled the positions of each PWM, then shuffled the nucleotides per position.

## Reference annotation based transcript assembly and estimation of mRNA levels using *D.melanogaster*, *D.yakuba*, *D.pseudoobscura* and *D.virilis* mRNAseq data

Reads were mapped to the genomes using tophat [59]. Refined annotations of *D. yakuba*, *D. pseudoobscura* and *D. virilis* were based on the reference annotations (Flybase releases 1.3, 2.22 and 1.2 respectively) and mRNA-seq data obtained from pools of embryos. The reference annotations and mapped reads were given as inputs to cufflinks (version 1.0.3) [42] with the options –g and '–3-overhang-tolerance 300'. Statistics on the new annotations are given in Table S4.

mRNA measurements were obtained as follows: reads from single-embryo mRNA-seq were mapped to the genome sequence using tophat and mRNA levels were estimated using cufflinks [60] and the above refined annotations (except *D. melanogaster* samples for which the reference annotation was used).

## Orthology assignment between genes of different species

Orthology assignment between genes was established based on the whole-genome alignment: genes were considered orthologous if the coordinates of their exons intersect more than 40% of their total length and if their orientation is the same (or unknown). Because this method is genome-alignment based, it takes into account both sequence similarity and synteny, thus favoring ortholog over paralog association. We removed from the analysis genes that showed orthology inconsistencies (e.g. genes with different orthologs in different species).

## Association between genes and regions bound by TFs

We associated TF binding and genes according to the following rule: we associated with each gene one of the 10,137

the merged regions (defined as the union of regions bound by any of the 4 TFs located within 10,000 bp from the gene and bound by the biggest number of TF over all four species. So a gene is always associated with a nearby region bound by 0 (if no TF bound within 10,000 bp from the gene) to all four TFs in any species.

## Reconstruction of ancestral values of TF binding and mRNA levels

Ancestral values were reconstructed according to a Brownian motion model [29] using the R package ape (function "ace"). Under a Brownian motion model, continuous characters evolve randomly following a random walk. The values at the three internal nodes, including the root, as well as values of binding variance were obtained for each cluster.

Quantitative divergence along the tree was estimated from the changes occurring along the six branches of the tree. It was calculated as the differences between the seven nodes of the tree (four leaves and three internal nodes). This estimation explicitly takes phylogenetic inertia into account. We also made six simpler pairwise comparisons between the four leaves, which do not take phylogenetic inertia into account, and thus do not make any assertion about phylogenetic processes.

Before comparing the variance of TF binding and mRNA levels according to a Brownian motion model (Figure 7C only), we scaled the normalized values with their standard deviation. We then used these scaled normalized values to reestimate maximum likelihood parameters of the model.

## Principal component analyses

The Principal component analyses were conducted on the concatenate of TF occupancies in the 4 species, so that the results from the different components would be comparable between species. Note that no axis seemed to be correlated with species-specific differences. As a control, we also performed a PCA on species-specific data and axes were very similar to the axes from the trans-species PCA, especially for the first axis. Correlation between TF occupancy values projected on PC1 from the global PCA and from species specific PCAs were above 0.99 for each species and were above 0.77 for the 3 other axes. The measurements of the contribution of each of the PC axes to within or between species TF variation were obtained from a multiple linear regression of TF binding on the different axes of the PCA.

## Linear regressions to predict TF binding and mRNA levels

We used multiple linear regression to model TF binding and TF binding divergence from associated factors such as TF-specific and Zelda motif enrichments, proximity to genes and the number of TFs bound to a locus. mRNA levels were modeled from associated BCD, GT, HB and KR binding.

To obtain the variance of TF binding explained by motif content, we fitted a linear regression of motif content for any $TFi$ at any locus $j$ in species $s$, using the R [61] function "lm" (default "stats" package) and the formula:

$$TFsij \sim motif\_based\_prediction\_sij$$

where $motif\_based\_prediction\_sij$ corresponds to binding at locus $j$ in species $s$ predicted solely on $TFi$ motif content [35].

We also predicted binding using more parameters (Figure S14A–D) including motif content. We fitted a multiple linear regression of these parameters for any $TFi$ using the

formula:

$$TFsij \sim motif\_based\_prediction\_sij + sum\_TFs\_binding\_locus\_j +$$

$$Zelda\_motif\_based\_prediction\_sj + Peak\_location\_j$$

where $sum\_TFs\_binding\_locus\_j$ corresponds to the total number of TFs binding this locus in all species. $Zelda\_motif\_based\_prediction\_sj$ corresponds to binding at locus $j$ predicted solely on Zld motif content. $Peak\_location\_j$ corresponds to the location of the peak in $D.$ $melanogaster$ (as described in Figure 5C).

Similarly, to obtain the variance of mRNA levels explained by the variance of TF binding intensity at the region associated with each gene, we fitted a multiple linear model of TF binding to mRNA levels using the function "lm" and the formula:

$$mRNAsj \sim BCDsj + GTsj + HRsj + KRsj.$$

where $mRNAsj$ corresponds to mRNA levels for gene $j$ in species $s$. The four other terms correspond to TF binding associated to gene $j$ in species $s$.

## Establishment of the list of A-P enhancers

We built the set of A-P enhancers used in Figure S9 from three sources: we included known A-P enhancers (as defined by [13]) and we screened by eye and included the regions from the RedFly database and the HOT regions [38] that drive the expression of a reporter gene along the A-P axis. The list of included regions is included as Table S5. Note that many regions overlap on the genome because the difference databases are redundant.

## Embryo collection and fixation for in situ hybridizations

Collection, fixation and prehybridization were performed as described in [62]. $D.$ $melanogaster$, $D.$ $yakuba$, $D.$ $pseudoobscura$ and $D.$ $virilis$ embryos were raised at $25°C$ for all species but $D.$ $virilis$ for which embryos were raised at $20°C$. Embryos were collected for 1 h30, 2 h, 2 h30 and 4 h and then aged for 1 h30, 1 h30, 2 h and 4 h, respectively. Embryos were then dechorionated in 50% bleach for 2 minutes, washed with milliQ water, and fixed for 25 minutes in a fixation solution (50% heptane, 4% formaldehyde, $0.5 \times$ PBS, 25 mM EGTA pH 8) with vigorous shaking. Embryos were devitellized by vigorously shaking in 1:1 heptane/cold methanol solution for 1 minute, and then rinsed three times with methanol and stored at $-20°C$.

## Probe synthesis and in situ hybridizations

Species-specific probes were designed to match the same gene part in each species. $Otd$ probe was $\sim 1000$ nt long and matched the 5′UTR part of the gene. $l(3)83Fd$ probe was $\sim 1200$ nt long and matched the last four exons of the gene. Finally $CG13894$ RNA probe was $\sim 1500$ nt long and matched the full-length gene.

RNA probes were synthesized as previously described with a few differences [63]. Probes were cloned by PCR amplification using cDNA prepared from mRNA extracted as described above. PCR products were closed into TOPO II and sequence verified. Probes were carbonate-treated. Primers used for cloning are listed in Table S6.

Before in situ hybridization, embryos were dehydrated in an ethanol series and soaked for one hour in xylene before three additional ethanol washes. Embryos were then post-fixed in three methanol washes and processed and mounted as described in [62]. Embryos were imaged and photographed using a Nikon Eclipse 80i microscope, with a Nikon DS-UI camera, and the NIS Elements F 2.20 software.

## Data availability

All reads will be made available at the NCBI GEO at the time of publication. Accession numbers and additional files are available at www.eisenlab.org/mparis. We thank Mark D. Biggin and Robert K. Bradley for their help in the initial phase of the project, as well as members of the Eisen lab for fruitful discussions and suggestions. We also thank the reviewers and the editor for their useful comments.

## Supporting Information

**Figure S1** Summary table of the dataset presented in this study. 1 to 3 ChIPs were performed for each of the TFs BCD, GT, HB and KR in each species among *D.melanogaster*, *D.yakuba*, *D.pseudoobscura* and *D.virilis*. Two types of antibodies were used (red: produced and purified using *D.melanogaster* epitopes; grey: produced using *D.melanogaster* epitopes and purified using *D.virilis* epitopes). *: experiments for which chromatin from *D. melanogaster*, *D. pseudoobscura* and *D. virilis* were pooled before ChIP. In addition, blastoderm embryos were processed for mRNAseq experiments. (PDF)

**Figure S2** Pooling chromatin before ChIP had virtually no effect on peak height measurements. From a GT ChIP experiment performed on a pool of *D. melanogaster*, *D. pseudoobscura* and *D. virilis* chromatins, we compared pooling effect on peak height measurements. Peak height was measured either using the >99% reads that mapped unambiguously to the genome sequence of one species only (x axis) *vs* <1% reads from that mapped to the genome sequence from several species, and from which we could not establish the species source (y axis). Points falling under the three oblique lines (plain, dashed and dotted) represent peaks for which pooling *may* have caused peak height measurement to be divided by at most half, 20% of 10%. (PDF)

**Figure S3** Qualitative and quantitative comparison of TF occupancy between replicates in (**A**) *D.melanogaster* (three replicates per TF) and (**B**) *D.pseudoobscura* (two replicates per TF), which are the 2 species for which replicates are available. **A**. For clarity on *D. melanogaster* plots, TF occupancy on the z axis was color-coded. (PDF)

**Figure S4** BCD, GT, HB and KR binding levels vary substantially between species. Pairwise comparison of raw BCD, GT, HB and KR binding measurements between species are shown between *D. melanogaster*, *D. yakuba*, *D. pseudoobscura*, *D. virilis*. (PDF)

**Figure S5** Overall sequence divergence is poorly correlated with binding divergence. The plots are similar to Figure 3A. Comparison of quantitative variation of BCD, GT, HB and KR binding divergence *vs.* underlying sequence divergence. Binding divergence was measured by the variance in a Brownian motion model of binding divergence, and sequence divergence was measured by the total length of a PhyML phylogenetic tree based on the underlying sequence alignment. (PDF)

**Figure S6** Enrichment of TF-specific motifs under called peaks for each ChIP. A 1 kb window centered on each peak summit was screened for the presence of TF-specific motifs, using PWMs from [22]. Motif enrichment over the window is displayed per position.

The red line corresponds to the same analysis using a randomized PWM. (PDF)

**Figure S7** TF-specific motifs are predictive of TF binding. Binding intensity was predicted in each cluster and each species based only from the enrichment of TF-specific motifs [35]. (PDF)

**Figure S8** Motif turnover is predictive of TF binding divergence. **A–D**. Binding divergence of BCD, GT, HB and KR along each branch of the tree is correlated with divergence of predicted TF binding, based only on TF-specific **E–H**. Same as **A–D**. Values were partitioned into three categories, depending on predicted changes of binding along a branch, based on binding motif turnover (thresholds indicated by vertical lines in **A–D**). ***: Wilcoxon test p-value<0.001. These plots are similar to Figures 4D and 4E. (PDF)

**Figure S9** TF binding falling within A-P enhancers regions is higher and better conserved than in the rest of the genome. A-P enhancers were defined as described in the method section and consist of regions that drive expression of a reporter gene along the A-P axis in early *D. melanogaster* embryos. Sets of orthologous peaks were partitioned into two categories, whether or not they intersect with A-P enhancers. Phylogenetic mean (**A**) and phylogenetic variance (**B**), were obtained from the Brownian motion model. **C**. Average number of species in for which binding was detected for each set of bound regions. *: Wilcoxon-test p-value<0.05 ; ***: Wilcoxon-test p-value<0.001. (PDF)

**Figure S10** Conclusions on differential binding conservation hold when using conditions less stringent than in the rest of the study for identifying sets of bound regions. Numbers of sets for each TF are indicated at the top of the figure, under TF name (for comparisons, main Figures were built from a total of 2061, 4191, 4986 and 5309 BCD, GT, HB and KR sets, respectively). **A**. Neighbor-joining trees based on pairwise distance matrices of TF occupancy at bound loci (Spearman's correlation coefficient was indicated). **B**. Proportion of the number of species for which TF was detected per cluster, from a species-specific peak ("1"), to a peak conserved in all 4 species ("4"). **C**. Comparison of qualitative conservation of TF binding in the different species. A conservation score, corresponding to the average number of species in which binding was detected (1–4), was calculated for each set of orthologous regions and ranked according to ancestral mean, as estimated using a Brownian motion model. **D**. Comparison of binding intensity, as represented by ancestral mean binding, and trans-species binding variance in a Brownian motion model of TF binding evolution. **E**. Mean binding conservation score (1–4 species) depending on peak location in *D. melanogaster*. **F**. Mean binding conservation score depending on the number of other A-P factors binding the same locus. To correct as much as possible for TF binding differences linked to different wiring sizes, clusters were binned into 10 bins, depending on the estimated ancestral values, and the conservation was estimated independently in each bin. The average conservation is displayed. All panels are similar to panels from Figures 3 and 5. (PDF)

**Figure S11** Conclusions on differential binding conservation hold when using conditions more stringent than in the rest of the study for identifying sets of bound regions. Numbers of sets for each TF are indicated at the top of the figure, under TF name (for comparisons, main Figures were built from a total of 2061, 4191,

4986 and 5309 BCD, GT, HB and KR sets, respectively). **A**. Neighbor-joining trees based on pairwise distance matrices of TF occupancy at bound loci (Spearman's correlation coefficient was indicated). **B**. Proportion of the number of species for which TF was detected per cluster, from a species-specific peak ("1"), to a peak conserved in all 4 species ("4"). **C**. Comparison of qualitative conservation of TF binding in the different species. A conservation score, corresponding to the average number of species in which binding was detected (1–4), was calculated for each set of orthologous regions and ranked according to ancestral mean, as estimated using a Brownian motion model. **D**. Comparison of binding intensity, as represented by ancestral mean binding, and trans-species binding variance in a Brownian motion model of TF binding evolution. **E**. Mean binding conservation score (1–4 species) depending on peak location in *D. melanogaster*. **F**. Mean binding conservation score depending on the number of other A-P factors binding the same locus. To correct as much as possible for TF binding differences linked to different wiring sizes, clusters were binned into 10 bins, depending on the estimated ancestral values, and the conservation was estimated independently in each bin. The average conservation is displayed. All panels are similar to panels from Figures 3 and 5.
(PDF)

**Figure S12** Principal Component Analysis of binding of all factors in each species. Each row represents a factor, and each column is a principal component of the relevant data. The color represents the sign (yellow positive, blue negative) and magnitude (color intensity) of each value in each principal component vector. In each case the sign of the first principal component is the same for all four factors, indicating that the dominant driver of both interspecies divergence and quantitative variation within single species is a coordinated change in binding strength of all factors.
(PDF)

**Figure S13** Zelda binding may drive BCD, GT, HB and KR binding in all four species. **A**. Zelda motif enrichment is highly predictive of Zelda binding in *D. melanogaster* Zelda binding was predicted based only from the presence of TF-specific motifs [35] and compared to measured Zelda binding in *D. melanogaster* blastoderm embryos [25]. **B**. Zelda binding predicted from motif enrichment is highly correlated with TF binding coordinates projected on PC1 in all four species.
(PDF)

**Figure S14** TF binding (**A–D**) and binding divergence (**E–L**) are better predicted by an integrative model of binding, rather than just motif enrichment. The binding (**A–D**), branch-wise binding divergence (**E–H**) and pairwise binding changes (**I–L**) of BCD (**A,E,I**), GT (**B,F,J**), HB (**C,G,K**) and KR (**D,H,L**) are well predicted by a multiple linear regression that takes into account motif enrichment, predicted Zelda binding, the nature of a nearby gene (if any) as well as the number of other TFs binding the same locus.
(PDF)

**Figure S15** mRNA levels are better predicted by associated nearby binding for zygotic than for maternal genes. **A–C**. Comparison of mRNA levels depending on the number of TFs associated with the gene. **D–F**. Comparison of mRNA levels between measured values and values predicted only on associated nearby TF binding.
(PDF)

**Figure S16** Divergence mRNA levels along the *Drosophila* tree are better predicted by associated divergence of nearby binding for zygotic than for maternal genes. **A–C**. Comparison of (**A–F**) branch-wise and (**G–I**) pairwise quantitative changes in mRNA levels depending on changes predicted mRNA levels based on associated TF binding. mRNA levels were predicted using a multiple linear regression. (**D–F**) Same as **A–C**. Values were partitioned into three categories, depending on predicted changes of binding along a branch. n.s: Wilcoxon test p-value>0.05 ; ***: Wilcoxon test p-value<0.05 ; **: Wilcoxon test p-value<0.01 ; ***: Wilcoxon test p-value<0.001.
(PDF)

**Figure S17** TF binding associated with zygotic genes is higher and better conserved than TF binding associated with maternal genes. TF binding divergence (**A**) and ancestral TF binding (**B**) for clusters associated with zygotic or maternal genes were calculated as in Figure 6. Binding divergence and ancestral values were estimated using a Brownian motion model of TF binding divergence. P-values of mean comparison (Wilcoxon test) are displayed above each graph.
(PDF)

**Figure S18** *In situ* hybridization of the gene *oc* that displays high divergence of nearby TF binding. **A**. Binding profile of BCD, GT, HB and KR for the four species, on the genomic alignment surrounding the gene. Gene limits are indicated at the top part of the panel, in addition to coordinates of sets of bound regions. Coordinates of the known enhancer "oc_otd_early_enhancer", whose activity mirrors *oc* expression pattern in *D. melanogaster* blastoderms (RedFly ID: RFRC:0000000373.004), is highlighted in yellow. Regions called as bound are highlighted under each profile. Of note, the region falling in the middle of the gene is not called as bound because the region did not pass our mappability filtering step in *D. virilis*. Coordinates refer to the genome alignment. **B**. *in situ* hybridizations of *oc* in the four studied species at four different developmental stages from early to late stage 5. Expression pattern in *D. melanogaster* is in agreement with the reported expression pattern from BDGP (http://insitu.fruitfly. org/cgi-bin/ex/report.pl?ftype = 1&ftext = CG12154).
(PDF)

**Figure S19** *In situ* hybridization of the gene *l(3)82Fd* that displays high divergence of nearby TF binding. **A**. Binding profile of BCD, GT, HB and KR for the four species, on the genomic alignment surrounding *l(3)82Fd*. Gene limits are indicated at the top part of the panel, in addition to coordinates of sets of bound regions. **B**. *in situ* hybridizations of *l(3)82Fd* in the four studied species at four different developmental stages from early to late stage 5. Expression pattern in *D. melanogaster* is in agreement with the reported expression pattern from BDGP (http://insitu.fruitfly. org/cgi-bin/ex/report.pl?ftype = 1&ftext = CG32464).
(PDF)

**Figure S20** *In situ* hybridization of the gene *CG13984* that displays high divergence of nearby TF binding. **A**. Binding profile of BCD, GT, HB and KR for the four species, on the genomic alignment surrounding *CG13984*. Gene limits are indicated at the top part of the panel, in addition to coordinates of sets of bound regions. . **B**. *in situ* hybridizations of *CG13984* in the four studied species at four different developmental stages from early to late stage 5. Expression pattern in *D. melanogaster* is in agreement with the reported expression pattern from BDGP (http://insitu.fruitfly. org/cgi-bin/ex/report.pl?ftype = 3&ftext = RE50383).
(PDF)

**Table S1** Collection conditions for each species (temperature and collection times).
(DOCX)

**Table S2** Mapping statistics for all ChIPs, including the number of mapped reads and the percentage of mapped reads that mapped uniquely to the genome, as a measure of library complexity.
(DOCX)

**Table S3**  Number of called peaks per TF per species.
(DOCX)

**Table S4**  Statistics on improvement of gene annotation using mRNA-seq. mRNA-seq data was used to improve reference annotations of *D. melanogaster*, *D. yakuba*, *D. pseudoobscura* and *D. virilis* using the RABT option of cufflinks. The increase in the number of bases covered by the annotation ad well as the number of new and modified genes and isoforms are indicated.
(DOCX)

**Table S5**  List of A-P enhancers used in this study (Figure S9). The list was obtained from three different and overlapping sources: known A-P target regions from [13], as well as regions from the RedFly database [39] and the HOT regions [38] that drive the expression of a reporter gene along the A-P axis in an early *D. melanogaster* embryo. Please note that the list was not corrected for redundancy and several regions may have overlapping coordinates.
(DOCX)

**Table S6**  Cloning primers for *in situ* hybridization.
(DOCX)

## Acknowledgments

We thank Mark D. Biggin and Robert K. Bradley for their help in the initial phase of the project, as well as members of the Eisen lab for fruitful discussions and suggestions. We also thank the reviewers and the editor for their useful comments.

## Author Contributions

Conceived and designed the experiments: MBE MP. Performed the experiments: MP XYL JEV SEL. Analyzed the data: MP TK. Contributed reagents/materials/analysis tools: TK. Wrote the paper: MBE MP.